\documentclass[conference]{IEEEtran}
\IEEEoverridecommandlockouts
\usepackage{cite}
\usepackage{amsmath,amssymb,amsfonts}
\usepackage{algorithmic}
\usepackage{graphicx}
\usepackage{url}
\usepackage{textcomp}
\usepackage{xcolor}
\usepackage{graphicx}
\usepackage{verbatim} 
\usepackage{multirow} 
\def\BibTeX{{\rm B\kern-.05em{\sc i\kern-.025em b}\kern-.08em
    T\kern-.1667em\lower.7ex\hbox{E}\kern-.125emX}}
\begin{document}

\title{Malware Evasion Attack and Defense}

\author{\IEEEauthorblockN{Yonghong Huang\IEEEauthorrefmark{1}, Utkarsh Verma\IEEEauthorrefmark{1}, Celeste Fralick\IEEEauthorrefmark{1}, Gabriel Infante-Lopez\IEEEauthorrefmark{2}, Brajesh Kumar\IEEEauthorrefmark{3}, Carl Woodward\IEEEauthorrefmark{1}}
\IEEEauthorblockA{
\textit{\IEEEauthorrefmark{1}McAfee, USA, \IEEEauthorrefmark{2}McAfee, Argentina, \IEEEauthorrefmark{3}McAfee, India}\\
\{yonghong\_huang, utkarsh\_verma1, celeste\_fralick, gabriel\_infante-lopez, brajesh\_kumar2, carl\_woodward\}@mcafee.com }
}

\maketitle

\begin{abstract}
Machine learning (ML) classifiers are vulnerable to adversarial examples. An adversarial example is an input sample which is slightly modified to induce misclassification in an ML classifier. In this work, we investigate \emph{white-box} and \emph{grey-box} evasion attacks to an ML-based malware detector and conduct performance evaluations in a real-world setting. We compare the defense approaches in mitigating the attacks. We propose a framework for deploying \emph{grey-box} and \emph{black-box} attacks to malware detection systems. 

\end{abstract}


\section{Introduction}
It has been shown that machine learning (ML) classifiers are vulnerable to adversarial examples. An adversarial example is a small perturbation of the original inputs and carefully crafted to misguide the classifier to produce incorrect output. As the presence of ML increases in malware detection ~\cite{Dahl2013, Saxe2015, Pascanu2015, Raff2017, Athiwaratkun2017, Anderson2018}, so does its likelihood of being attacked, hijacked or manipulated. In this work, we examine the potential vulnerabilities of ML-based malware detectors in adversarial environments. We hope that our findings may motivate better use and safe practices of ML for security applications. 

The study of adversarial examples has continued to grow. Biggio provided an overview of adversarial ML for the past 10 years ~\cite{Biggio2018}. There have been extensive studies of adversarial examples in computer vision ~\cite{Szegedy2013, Goodfellow2015, Papernot2016, Moosavi-Dezfooli2016, Madry2017, Carlini2017, Athalye2018}. Szegedy ~\cite{Szegedy2013} first identified adversarial examples in deep neural networks (DNN) and proposed L-BFGS attack. Goodfellow  ~\cite{Goodfellow2015} proposed the Fast Gradient Sign Method (FGSM), which is a simple and fast method to generate adversarial examples for practical adversarial training. Carlini ~\cite{Carlini2017} proposed the C\&W attacks by minimizing the $L0$, $L2$ and $L\infty$ distance metrics. It is one of the strongest attacks. Papernot ~\cite{Papernot2016} proposed the Jacobian-based Saliency Map Approach (JSMA) which minimize the number of features and magnitude to be manipulated. Athalye~\cite{Athalye2018} proposed the expectation over transformation (EOT) algorithm to generate adversarial examples which remain adversarial over a chosen transformation distribution. 

Recent research of adversarial examples in malware detection has drawn attention in both ML and security communities ~\cite{Grosse2017, Kolosnjaji2018, Kreuk2019, Demetrio2019}. Grosse ~\cite{Grosse2017} presented the adversarial examples for Android malware. Demetrio~\cite{Demetrio2019} used an interpretable ML approach to identify the most influential features and proposed an attack to generate adversarial malware examples by manipulating the file header. There are some work on \emph{black-box attack} on images ~\cite{Liu2016, Papernot2017} and malware ~\cite{Hu2017}. However, they are not in real-world black-box setting. 

Traditional techniques for making ML robust, such as weight decay or dropout, do not provide a practical defense. Adversarial training ~\cite{Goodfellow2015} and defensive distillation ~\cite{Papernot22016} are two significant defense methods. Carlini ~\cite{Carlini22017} shows that the current defense methods including ~\cite{Goodfellow2015, Papernot22016} are not effective. 

The goal of this work is to study the vulnerabilities of an ML-based malware detector and develop a solution resilient to the evasion attack. The main contributions are

\begin{itemize} 
	\item {We conducted \emph{white-box} \& \emph{grey-box} attacks to an ML-based malware detector with thorough evaluations.}	
	\item {We developed defense mechanisms and compared their effectiveness.}  
	\item {We propose a framework for deploying \emph{gray-box} and \emph{black-box} attacks in a real-world setting.}    
\end{itemize}

\section{Methodology} 
\label{section:Methodology}

\subsection{Data Description}
\label{section:DataDescription}

The training data were collected by McAfee Labs in January and February 2018. A subset of the training samples and the trained DNN model (trained with millions of samples) were provided. Table ~\ref{tab:datasets} shows the dataset including training set, validation set and testing set. The testing data were collected from VirusTotal, which are independent of the training data. 

The $491$ API features were extracted from log files. Table ~\ref{tab:log} shows an excerpt of a log file. The logs capture the API calls from the source files. Portable executable (PE) samples (exe and dll) were used to generate logs. The mixed data, which contained API logs generated from Win7, WinXP, Win8, and Win10, were created. The raw counts of the APIs were applied to feature transformation and the values were normalized to [0,1]. Table ~\ref{tab:apis} shows an excerpt of the API features. 

\begin{table}[t]
	\caption{The Dataset}
	\center
	\label{tab:datasets}
	\small
	\begin{tabular}{{|c|c|}}
		\hline
		\textbf{Dataset} &\textbf{Number of Samples}\\\hline
		Training Set     & 57170 (28594 clean and 28576 malware)  \\\hline
		Validation Set     & 578 (280 clean and 298 malware)   \\\hline		
		Test Set     & 45028 (16154 clean and 28874 malware)   \\\hline 
			
	\end{tabular}
\end{table}

\begin{table}[t]
	\caption{Excerpt of a log file}
	\center
	\label{tab:log}
	\small
	\scriptsize\tt
	\begin{tabular}{|l|c|}
		\hline		
		GetStartupInfoW:7FEFDD39C37 ()"61468" \\
		GetFileType:7FEFDD39D0C ()"61468" \\  
		GetModuleHandleW:13FBC34C3 ()"61484" \\
		GetProcAddress:13FBC34D6 (76D30000,"FlsAlloc")"61484" \\
		...\\
		GetStartupInfoW:13FBC4539 ()"61484" \\
		GetStdHandle:13FBC46F1 ()"61484" \\
		GetFileType:13FBC4707 ()"61484" \\
		FreeEnvironmentStringsW:13FBC4D49 ()"61484" \\
		GetCPInfo:13FBC263D ()"61484" \\
		\hline
	\end{tabular}
\end{table}

\begin{table}[t]
	\caption{Excerpt of the API features}
	\center
	\label{tab:apis}
	\small
	\scriptsize\tt
	\begin{tabular}{|l|c|}
		\hline		
		475 waitmessage \\
		476 windowfromdc \\
		477 winexec \\
		478 writeconsolea \\
		479 writeconsolew \\
		480 writefile \\
		481 writeprivateprofilestringa \\
		482 writeprivateprofilestringw \\
		483 writeprocessmemory \\
		484 writeprofilestringa \\		
		\hline
	\end{tabular}
\end{table}

\subsection{Attack Experimental Setup}
\label{subsection:ExperimentalSetup}
We design the \emph{white-box}, \emph{grey-box} and \emph{black-box} attack experiments. In \emph{white-box attack}, the attacker has complete knowledge of the system, including training data, features, and ML models (i.e. DNN architecture and parameters). In \emph{grey-box attack}, the attacker has no knowledge of training data and ML model, but knowledge of the features. In \emph{black-box attack}, the attacker has no knowledge of the system. 

\subsubsection{JSMA for Adversarial Examples}
\label{subsubsection:JSMAattack}

In this work, we aim to develop an attack with the perturbations using minimum number of features, so we selected the JSMA ~\cite{Papernot2016}. The success rate and transfer rate for the JSMA are relatively high.

The JSMA allows us to select the most important feature associated with the maximum gradient based on the saliency map, and perturb the feature to increase the likelihood of the input as the target class. We compute the gradient of $F$ with respect to $X$ to estimate the direction in which a perturbation in $X$ would change $F$’s output. A perturbation of $X$ with maximal positive gradient into the target class $0$ (clean) is chosen. For $F(X) = F_0(X), F_1(X)$, 

\begin{equation}
\bigtriangledown_{F}=\frac{\partial{F(X)}}{\partial{X}}=[\frac{\partial{F_i(X)}}{\partial{X_j}}]_{i\in{0,1}, j\in{[1,M]}}, 
\end{equation}

\begin{figure}[!t]
	\centering
	\includegraphics[width=3.3in]{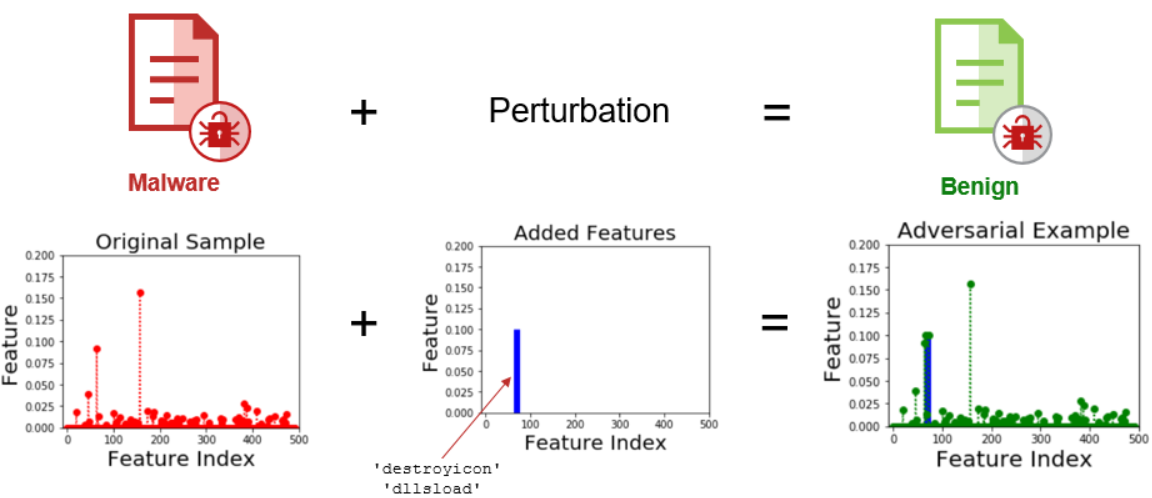}
	\caption{Generate an adversarial example for malware. The left is one malware sample with 491 features. The middle one is adding two API calls - 'destroyicon' and 'dllsload'. The right is the adversarial example.}
	\label{fig:advExample}
\end{figure}

Where $M$ is the number of features; $i =0, 1$ is clean and malware. We ensure that only API calls are added and not deleting any existing features. We make sure the functionality of malware unchanged. CleverHans ~\cite{Papernot2018} was used for the JSMA implementation. The $\theta$ controls the perturbations introduced to modified features. The $\gamma$ controls the maximum percentage of perturbed features, which is associated with the number of perturbed features. Figure ~\ref{fig:advExample} shows how to generate an adversarial example for malware. The adversarial example evades the malware detector and is recognized as benign. 

\subsubsection{Transferability of Adversarial Examples}

\begin{figure}[!t]
	\centering
	\includegraphics[width=3in]{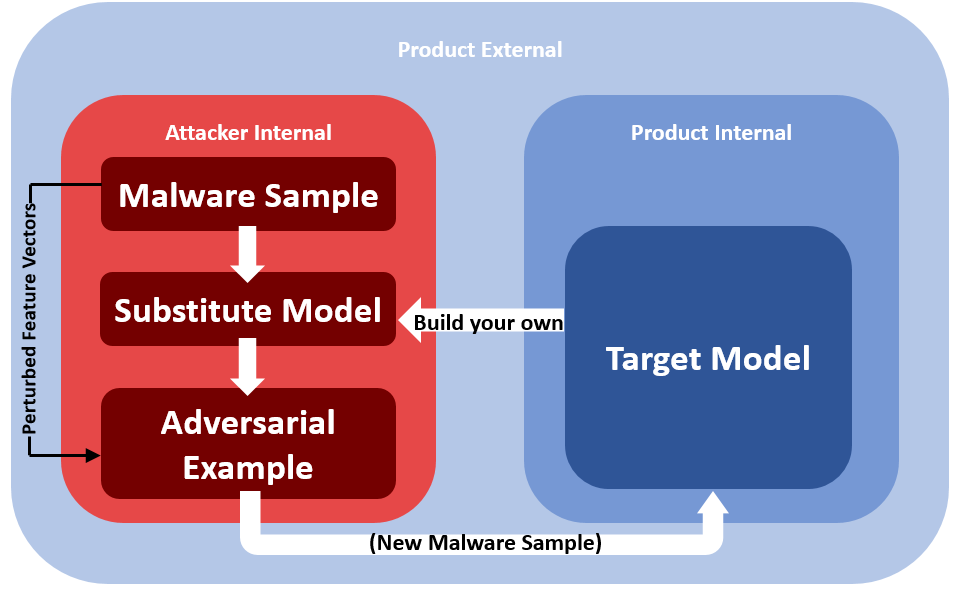}
	\caption{A framework for \emph{grey-box} and \emph{black-box} attacks in real-world setting}
	\label{fig:transferability}
\end{figure}

Adversarial examples generated from one ML model can be used to misclassify another model, even if both models have different architectures and training data. This property is called transferability. This property holds because a contiguous subspace with a large dimension in the adversarial space is shared among different models ~\cite{Tramr22017}. The transferability of adversarial examples makes the \emph{grey-box} and \emph{black-box} attacks become possible. Papernot~\cite{Papernot2017} shows that multiple ML algorithms are vulnerable due to the transferability of adversarial examples. Figure ~\ref{fig:transferability} shows the framework for \emph{grey-box} and \emph{black-box} attacks in real-world testing and the transferability of adversarial examples in malware detection. The attacker can train a substitute model to craft adversarial examples and deploy them to the target model to evade the detection. The attacker's model, features and training data are different from the target model, features and training data. In this work, the target model is 4-layer fully connected DNN (The target model is proprietary, so we cannot release the detail information.) Table ~\ref{tab:attacks} summarizes the architectures of the substitute model, an 5-layer DNN. 

\begin{table}[t]
	\caption{The substitue model}  
	\center
	\label{tab:attacks} 
	\small
	\begin{tabular}{{|l||l|}}
		\hline		
		57170 balanced training data  \\\hline
		5-layer DNN \\\hline		
		1st layer – 491 nodes \\
		2nd layer – 1200 nodes \\
		3rd layer – 1500 nodes \\
		4th  layer – 1300 nodes \\
		5th layer – 2 nodes    \\
		\hline
	\end{tabular}
\end{table}

\subsection{Defense Methods}
\label{subsection:DefenseMechanisms}
We applied the four defense approaches for the ML malware engine. Our considerations in defense are low impact on model architecture and model speed, and maintain model accuracy.  

\subsubsection{Adversarial Training}
The basic idea of adversarial training ~\cite{Szegedy2013, Goodfellow2015} is to inject adversarial examples into training process and train the model so that the model can generalize to defense against the attacks. The approach has been shown effective with only a small loss of accuracy. However, it needs adversarial examples to train the model, and the defense performance decreases for different attack methods.  

\subsubsection{Defensive Distillation}
There are two models in defensive distillation ~\cite{Papernot22016}. The first model is trained as usual. The second model (compressed model) is trained with the probabilities (soft labels) learned from the first one. The benefit of using soft class labels lies in the additional knowledge in probabilities, compared to hard class labels. Defensive distillation prevents models from fitting too tightly to the data and contributes to better generalization. Distillation is controlled by the softmax temperature $T$. High $T$ force DNN to produce probabilities with large values for each class.

\subsubsection{Feature Squeezing}
Feature squeezing ~\cite{Xu2017} reduces the degrees of freedom for the attacker to construct adversarial examples by squeezing out unnecessary input features. We used $L1$ norm to measure the distance between the model's prediction on the original sample and the prediction on the sample after squeezing. If the distance is larger than a threshold, then the input sample is an adversarial example. Otherwise, it is a legitimate sample. The assumption is that the squeezing significantly impacts the classifier's prediction on the adversarial examples while it has no significant impact on the classifier's prediction on the legitimate samples. 

\subsubsection{Dimensionality Reduction}
The approach uses Principal Component Analysis (PCA) for dimensionality reduction ~\cite{Bhagoji2017}. Instead of training a classifier on the original data, it reduces the features from the $n$-dimension to $k$ ($k<<n$), and trains the classifier on the reduced input. The defense restricts the attacker to the first $k$ components. The assumption is that adversarial examples rely on principal components. Therefore, restricting the attack to the first $k$ components should increase the required distortion to produce adversarial examples. 

\subsection{Evaluation Metrics}
For attack evaluation, we used the security evaluation curve (detection rate as a function of attack strength), transferability in terms of transfer rate, and perturbation measured by $L_{2}$ distance norm between the adversarial examples and the original examples. For defense evaluation, we use confusion matrix - true positive rate (TPR), true negative rate (TNR), false positive rate (FPR), and false negative rate (FNR). 

\section{Results and Discussion}
\label{section:Results} 

\subsection{White-box Attack}

\begin{figure}[htb]
	
	\begin{minipage}[b]{.48\linewidth}
		\centering
		\centerline{\includegraphics[width=1.6in]{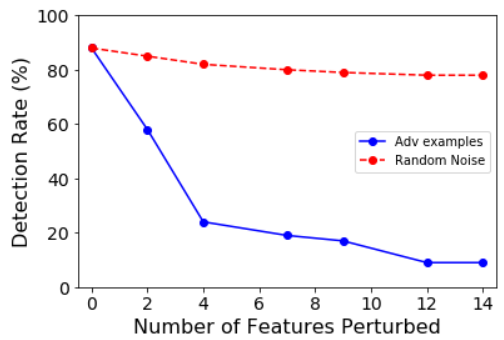}}
		\centerline{(a)$\theta = 0.100$}\medskip
	\end{minipage}
	\hfill
	\begin{minipage}[b]{0.48\linewidth}
		\centering
		\centerline{\includegraphics[width=1.6in]{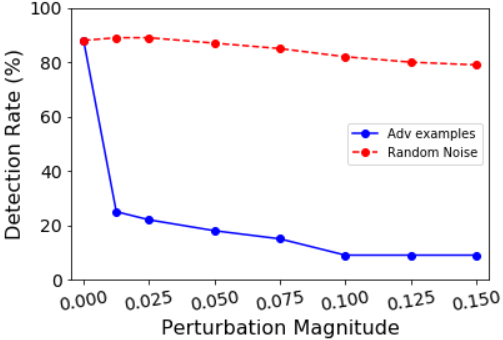}}
		\centerline{(b)$\gamma =0.025$}\medskip
	\end{minipage}
	\caption{Security evaluation curve for \emph{white-box} attack. $\gamma$ is assciated with the number of perturbed features. $\theta$ is the magnitude of perturbed features.}
	\label{fig:whitebox}
\end{figure}

\begin{figure}[htb]
	
	\begin{minipage}[b]{0.48\linewidth}
		\centering
		\centerline{\includegraphics[width=1.6in]{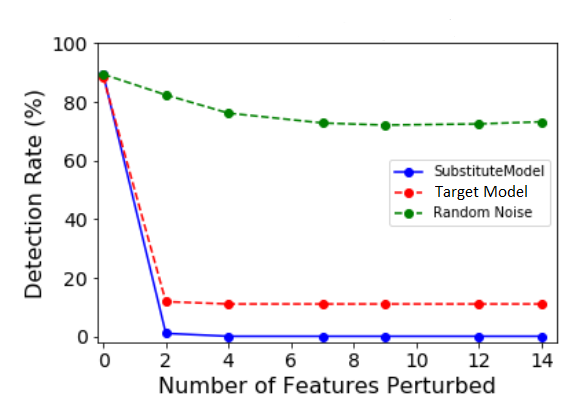}}
		\centerline{(a)$\theta = 0.100$}\medskip
	\end{minipage}
	\hfill
	\begin{minipage}[b]{0.48\linewidth}
		\centering
		\centerline{\includegraphics[width=1.6in]{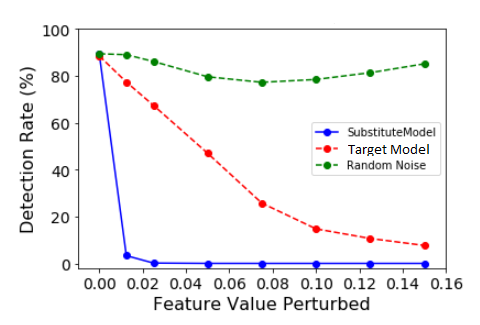}}
		\centerline{(b)$\gamma =0.005$}\medskip
	\end{minipage}
	\hfill
	\begin{minipage}[b]{0.48\linewidth}
		\centering
		\centerline{\includegraphics[width=1.6in]{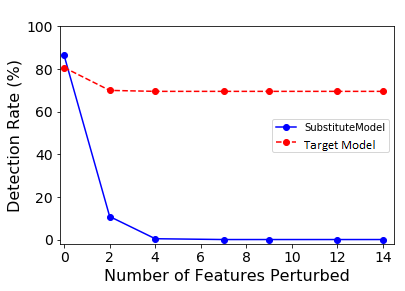}}
		\centerline{(c)$\theta = 0.100$, binary features}\medskip
	\end{minipage}
	\caption{Security evaluation curves for \emph{gray-box} attacks. $\gamma$ is the number of perturbed features and $\theta$ is the magnitude of perturbed features.}
	\label{fig:attacks}
\end{figure}

\begin{figure}[htb]
	
	\begin{minipage}[b]{.48\linewidth}
		\centering
		\centerline{\includegraphics[width=1.6in]{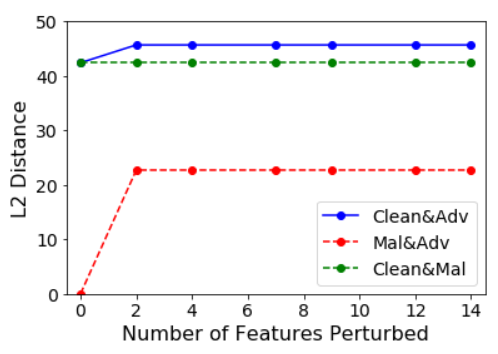}}
		\centerline{(a)$\theta = 0.100$}\medskip
	\end{minipage}
	\hfill
	\begin{minipage}[b]{0.48\linewidth}
		\centering
		\centerline{\includegraphics[width=1.6in]{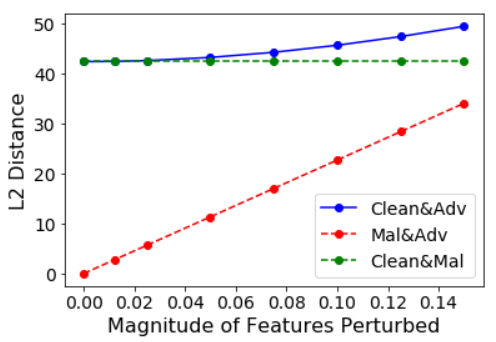}}
		\centerline{(b)$\gamma =0.005$}\medskip
	\end{minipage}
	\caption{L2 distances in \emph{gray-box} attack using original features}
	\label{fig:L2}
\end{figure}

The \emph{white-box} attack on $28,874$ malware samples is shown in Figure ~\ref{fig:whitebox}, where (a) $\theta = 0.1$ and $\gamma = [0:0.005:0.030]$ adding $[0:2:14]$ features, and (b) $\gamma = 0.025$ and $\theta = [0:0.0125:0.15]$. For both (a) and (b), the detection rates drop significantly when attack strength increases either by adding more API calls or increasing the frequency of the API calls. Randomly adding features does not decrease the detection rates. The results show that the \emph{white-box} evasion attack is effective. The JSMA perturbation is different from random noise. In the operating point $\theta = 0.1$ and $\gamma = 0.025$ (adding $12$ features), the detection rate drops to 0.099, which indicates $26,015$ malware evasion. 

\subsection{Grey-box Attack}

We design three experiments for \emph{gray-box} attacks. For the first one, we assume the attacker knows the exact 491 features. The substitute model is used to craft the adversarial examples, which are then deployed to the target model. We train the substitute model with $1000$ epoch, batch size = $256$ and learning rate = $0.001$ and Adam optimizer. We use $28,874$ malware samples to craft adversarial examples. The \emph{gray-box} attack is shown in Figure ~\ref{fig:attacks}, where (a)$\theta = 0.1$ and (b)$\gamma =0.005$ (i.e. adding $2$ features). The results show that the \emph{gray-box} attack is effective. The detection rates of the target model significantly decrease with a small number of perturbed features or with small amount magnitude changes. It indicates that the attacker can craft the adversarial examples using the substitute model to evade the DNN malware detector. In the operating point $\theta = 0.1$ and $\gamma = 0.005$, which associated with adding $2$ features, the detection rate of the target model drops to 0.147. The transfer rate is $0.853$. It indicates that $24,630$ malware can evade the malware detector. The \emph{gray-box} attack is less effective than the white-box attack because the attacker has less knowledge about the target system. 

For the second experiment, we assume the attacker does not have knowledge of the DNN model and the feature transformation but knows API calls as features. We create a substitute model using binary features - when the API appears, the feature value equals one, otherwise equals to zero. Figure \ref{fig:attacks} (c) shows the detection rates decrease significantly when the attack strength increases by adding more API calls for the substitute model, but does not significantly impact the target model. The detection rate of the target model is $0.6951$. The transfer rate is $0.3049$. It shows the attack is less effective when attacker has less knowledge about the features. 

To understand adversarial examples across the decision boundary, we evaluated the L2 distances for (1) malware and adversarial examples (2) malware and clean (3) clean and adversarial examples. The L2 results are shown in Figure ~\ref{fig:L2}, where (a)$\theta = 0.1$ and (b)$\gamma =0.005$ (i.e. adding $2$ features). Both plots show consistent results. The results show that the distance for malware and adversarial examples is the shortest and the distance for clean to adverarial examples is the largest. The distance increases when the attack strength increase. From the results, we can see that the adversarial examples in high dimension feature space are in the blind spot far away from clean, and not lie in the decision boundary between malware and clean,  which provides us the insight for defense. 

For the third experiment, we conducted the live \emph{gray-box} testing. We were provided a source file and an associated log file. We used the substitute model to generate an adversarial example. We asked a security researcher to add one single API call multiple times in the source code and ran the DNN engine to detect this sample. The DNN engine originally detects this sample as malware with $98.43\%$ confidence. After adding this API once in the source code, the detection rate drops to $88.88\%$ confidence. When adding the same API eight times in the source code, the detection rate drops to $0\%$. It indicates the adversarial example successfully evaded the DNN malware detector in a real-world \emph{gray-box} setting. 

\subsection{Defense}

\begin{table}[t]
	\caption{Adversarial Training Dataset}
	\center
	\label{tab:AdvTrainingData1}
	\small
	\begin{tabular}{{|c|c|}}
		\hline
		\textbf{Dataset } &\textbf{Number of Samples}\\\hline
		Training Set     & 53482(26118 clean, 27364 malware and advEx) \\\hline			
		Test Set         & 26560(5090 clean, 5252 malware and 16218 advEx) \\\hline 
		
	\end{tabular}
\end{table}

\begin{table}[t]
	\caption{Defense Testing Results}
	\center
	\label{tab:defenseResult}
	\small
	\begin{tabular}{|c|c|c|c|c|c|}
		\hline
		\textbf{}          &\textbf{Dataset Name} &\textbf{TPR} &\textbf{TNR} \\\hline
		\multirow{ 3}{*}{No Defense} & Clean Test        &   nan       & 0.964      \\ 		
		& Malware Test       &   0.883    & nan         \\		
		& AdvExamples  &   0.304    & nan         \\\hline
		\multirow{ 3}{*}{AdvTraining}  & Clean Test         &   nan      & 0.995           \\ 		
		& Malware Test       &   0.888    &  nan        \\		
		& AdvExamples  &   0.931    &  nan         \\\hline
		\multirow{ 3}{*}{Distillation}  & Clean Test         &   nan      & 0.428         \\ 		
		& Malware Test       &   0.573      & nan          \\		
		& AdvExamples  &  0.577   & nan        \\\hline
		\multirow{ 3}{*}{FeaSqueezing}  & Clean Test         &   nan     & 0.586        \\ 		
		& Malware Test       &   nan    & 0.438        \\		
		& AdvExamples  &   0.554   & nan         \\\hline
		\multirow{ 3}{*}{DimReduct}  & Clean Test         &   nan      & 0.674       \\ 		
		& Malware Test       &   0.914    & nan      \\		
		& AdvExamples  &   0.913    &  nan      \\\hline
	\end{tabular}
\end{table}

We applied the four defense methods in Section ~\ref{subsection:DefenseMechanisms}. The results of the defense approaches are shown in Table \ref{tab:defenseResult}.  In adversarial training, we did sanity check on the data to reduce the duplicated samples. A subset of the adversarial examples from the \emph{gray-box} attack ($\theta = 0.1$ and $\gamma = 0.02$) and a subset of testing malware were added to the training set. In order to make the training set balance, we added a subset of clean samples into the training set. Table \ref{tab:AdvTrainingData1} shows the dataset. The testing results show that adversarial training significantly increases the detection rate for adversarial examples from $0.304$ to $0.931$. At the same time, it also increases the TNR from $0.964$ to $0.998$ without compromising the detection rate for the original malware. In dimension reduction, we selected $K=19$. The testing results show that both detection rates of adversarial examples and malware significantly increases to $0.913$ and $0.914$. However, the TNR decreases from $0.964$ to $0.674$. The results suggest we may consider ensemble adversarial training and dimension reduction. For both defensive distillation ($T= 50$) and feature squeezing, the detection rate of adversarial examples increases. However, the TNR for detecting clean and TPR for detecting malware drop. 

\section{Conclusion}
\label{section:Conclusion}

The JSMA attack shows that modifying one bit in the feature vector can bypass the real-world ML-based malware detector. The attack can be done by modifying the malware source code by adding one API call. We show some preliminary results on building attack resiliency to API additions for mitigation. It is an open challenge to design a defense against a powerful adaptive attack. For future work, we are building the real-world \emph{black-box} testing framework as proposed in Figure ~\ref{fig:transferability} using open source data with different features and models. We will study the intepretability of adversarial examples to develop more effective defenses.

\section*{Acknowledgment}

We would like to thank Vaisakh Shaj and Xiang Xu for their engineering contributions. Additional thanks to Esteban Prospero, Muralivardhan Pannala, Sidney Gomindes for their support to the project.


\begin{thebibliography}{9}
	
	\bibitem{Dahl2013}
	G. E. Dahl, J. W. Stokes, L. Deng, and D. Yu, \emph{Large-scale malware classification using random projections and neural networks}, in Proc of IEEE Intl. Conf on Acoustics, Speech and Signal Processing (ICASSP), 2013.
	
	\bibitem{Saxe2015}
	J. Saxe and K. Berlin, \emph{Deep neural network based malware detection using two dimensional binary program features}, arXiv preprint arXiv:1508.03096, 2015.
	
	\bibitem{Pascanu2015}
	R. Pascanu, J. W. Stokes, H. Sanossian, M. Marinescu, and A. Thomas, \emph{Malware classification with recurrent networks}, in Proc of ICASSP, 2015.
	
	\bibitem{Raff2017}
	E. Raff, J. Barker, J. Sylvester, R. Brandon, B. Catanzaro, and C. Nicholas, \emph{Malware detection by eating a whole exe}, arXiv preprint arXiv:1710.09435, 2017.
	
	\bibitem{Athiwaratkun2017}
	B. Athiwaratkun and J.W. Stokes, \emph{Malware classification with LSTM and GRU language models and a character-level CNN}, in Proc of ICASSP, 2017.
	
	\bibitem{Anderson2018}
	H. S. Anderson and P. Roth, \emph{EMBER: An Open Dataset for Training Static PE Malware Machine Learning Models}, arXiv preprint arXiv:1804.04637, 2018.
	
	\bibitem{Biggio2018}
	B. Biggio and F. Roli, \emph{Wild Patterns: Ten Years After the Rise of Adversarial Machine Learning}, arXiv preprint arXiv:1712.03141, 2018.
		
	\bibitem{Szegedy2013}
	C. Szegedy,  W. Zaremba, I. Sutskever, D. Erhan, I. Goodfellow and R. Fengus, \emph{Intriguing properties of neural networks}, in Proc of Intl Conf on Learning Representations (ICLR), 2013.
	
	\bibitem{Goodfellow2015}
	I.J. Goodfellow, J. Shlens, and C. Szegedy, \emph{Explaining and harnessing adversarial examples}, in Proc of ICLR, 2015.
	
	\bibitem{Papernot2016}
	N. Papernot, P. McDaniel, S. Jha, M. Fredrikson, Z. B. Celik, and A. Swami, \emph{The limitations of deep learning in adversarial settings}, in Proc of IEEE European Symposium on Security \& Privacy, 2016. 
	
	\bibitem{Carlini2017}
	N. Carlini and D. Wagner, \emph{Towards evaluating the robustness of neural networks}, in Proc of IEEE Symposium on Security and Privacy, 2017. 
		
	\bibitem{Athalye2018}
	A. Athalye, L. Engstrom, A. Ilyas, K. Kwok, \emph{Synthesizing Robust Adversarial Examples}, in Proc of Intl Conf on Machine Learning (ICML), 2018. 
	
	\bibitem{Moosavi-Dezfooli2016}
	S. Moosavi-Dezfooli, A. Fawzi and P. Frossard, \emph{Synthesizing robust adversarial examples}, in Proc of IEEE Conf on Computer Vision and Pattern REcognition (CVPR), 2016.
		
	\bibitem{Madry2017}
	A. Madry. A. Makelov, L. Schmidt, D. Tsipra and A. Vladu, \emph{Towards Deep Learning Models Resistant to Adversarial Attacks}, arXiv preprint arXiv:1706.06083, 2017.
	                      
	\bibitem{Papernot2018}
	N. Papernot et al., \emph{Technical report on the CleverHans v2.1.0 adversarial examples library}, arXiv preprint arXiv:1610.00768, 2018. 
		
	\bibitem{Grosse2017}
	K. Grosse, N. Papernot, P. Manoharan, M. Backes, P. D. McDaniel, \emph{Adversarial examples for malware detection}, ESORICS (2), Vol. 10493 of LNCS, Springer, pp. 62–79, 2017.
	
	\bibitem{Kolosnjaji2018}
	B. Kolosnjaji, A. Demontis, B. Biggio, D. Maiorca, G. Giacinto, C. Eckert, and F. Roli, \emph{Adversarial malware binaries: Evading deep learning for malware detection in executables}, arXiv preprint arXiv:1803.04173, 2018.
	
	\bibitem{Kreuk2019}
	F. Kreuk, A. Barak, S. Aviv, M. Baruch, B. Pinkas, and J. Keshet., \emph{Deceiving end-to-end deep learning malware detectors using adversarial examples}, arXiv preprint arXiv:1802.04528, 2019.
	
	\bibitem{Demetrio2019}
	L. Demetrio, B. Biggio, G. Lagorio,  F. Roli and A. Armando, \emph{Explaining Vulnerabilities of Deep Learning to Adversarial Malware Binaries}, arXiv preprint arXiv:1901.03583, 2019.	
	
	\bibitem{Liu2016}
	Y. Liu, X. Chen, C. Liu, and D. Song, \emph{Delving into transferable adversarial examples and black-box attacks}, arXiv preprint arXiv:1611.02770, 2016.
	
	\bibitem{Papernot2017}
	N. Papernot, P. McDaniel, I. Goodfellow, S. Jha, Z. B. Celik, A. Swami, \emph{Practical black-Box attacks against machine learning}, in Proc of ACM on Asia Conference on Computer and Communications Security, 2017.
	
	\bibitem{Hu2017}
	W. Hu and Y. Tan, \emph{Generating adversarial malware examples for black-Box attacks based on GAN}, arXiv preprint arXiv:1702.05983, 2017.
	
	\bibitem{Papernot22016}
	N. Papernot, P. McDaniel, X. Wu, S. Iha and A. Swami, \emph{Distillation as a defense to adversarial perturbations against deep neural networks}, In Proc of IEEE Symposium on Security \& Privacy, 2016.
	
	\bibitem{Carlini22017}
	N. Carlini and D. Wagner, \emph{Adversarial examples are not easily detected: bypassing ten detection methods}, arXiv preprint arXiv:1705.07263v2, 2017.
	
	\bibitem{Xu2017}
	W. Xu, D. Evans, and Y. Qi, \emph{Feature squeezing: Detecting adversarial examples in deep neural networks}, In Proc of Network and Distributed Systems Security Symposium, 2018. 
	
	\bibitem{Tramr22017}
	F. Tramr, N. Papernot, I. Goodfellow, D. Boneh, P. McDaniel, \emph{The space of transferable adversarial examples},  arXiv preprint arxiv:1704.03453, 2017. 
	
	\bibitem{Bhagoji2017}
	A. N. Bhagoji, d. Cullina, C. Sitawarinn, P. Mittal, \emph{Enhancing robustness of machine learning systems via data transformations}, arXiv preprint arxiv:1704.02654v4, 2017.
	
\end{thebibliography}
\end{document}